\documentclass{iopart}

\usepackage{amsfonts}
\usepackage{graphicx} 
\usepackage{times}
\usepackage{geometry} 
\usepackage[sort&compress]{natbib} 
\usepackage{bm} 
\newcommand{\flatd}{\partial}
\newcommand{\laplacian}{\nabla^2}

\newcommand{\eqref}[1]{(\ref{#1})}
\begin{document}
\pacs{04.20.Ex, 04.25.D-, 04.25.dg, 04.25.dc}

\title{Distributional sources for black hole initial data}

\author{Aaryn Tonita}
\address{Max-Planck-Institut f{\"u}r Gravitationsphysik, Albert-Einstein-Institut, Potsdam-Golm, Germany}
\address{Universit{\" a}t Potsdam, Potsdam, Germany}
\ead{atonita@aei.mpg.de}

\begin{abstract}
Black hole initial data is usually produced using Bowen-York type puncture initial data or by applying an excision boundary condition. The benefits of the Bowen-York initial data are the ability to specify the spin and momentum of the system as parameters of the initial data. In an attempt to extend these benefits to other formulations of the Einstein constraints, the puncture method is reformulated using distributions as source terms. It is shown how the Bowen-York puncture black hole initial data and the trumpet variation is generated by distributional sources. A heuristic argument is presented to argue that these sources are the general sources of spin and momentum. In order to clarify the meaning of other distributional sources, an exact family of initial data with generalized sources to the Hamiltonian constraint are studied; spinning trumpet black hole initial data and black hole initial data with higher order momentum sources are also studied.
\end{abstract}

\section{Introduction}
\label{sec:introduction}
The necessity that gravitational collapse leads to the generic formation of singularities distinguishes general relativity from other classical theories where point sources only enter as idealizations. Unfortunately, the use of distributions to describe sources in Einstein's equations is limited to a subset of distributions which is just regular enough for the equations to be well defined as shown in \cite{ads:1987PhRvD..36.1017G}. These distributions must not be too singular and have at most codimension 1. Nevertheless, many studies have described the structure of various singular spacetimes using linear distribution theory or generalized function algebras; these studies are reviewed in \cite{ads:2006CQGra..23R..91S}. In particular, \cite{ads:1979JMP....20.1423P} characterized the stress energy of the Schwarzschild spacetime in Kerr-Schild coordinates using Schwartz distributions while the Kerr solution has been characterized using Colombeau algebras(a differential algebra which seeks to provide a rigorous multiplication of generalized functions) in \cite{ads:1979JMP....20.1423P}.

In spite of the difficulties involved with the use of distributions in general relativity, distributions have been implicitly used in the construction of initial data for quite some time; this will be shown in detail in this paper. In particular, the puncture family of initial data (including the trumpet type initial data) can be interpreted as arising from differential equations with distributional sources. The usual derivation of puncture initial data considers that the data exists only on a punctured plane $\mathbb{R}^3 - \{\bm{x}_I\}$ with $\{\bm{x}_I\}$ the locations of an arbitrary number of punctures; these locations are interpretable as an asymptotically flat extended sheet of the spacetime (see \cite{ads:1997PhRvL..78.3606B}) and therefore as coordinate singularities. Boundary conditions are not imposed at these inner boundaries, which leads to an essential non-uniqueness to the constraint equations: considering the linear part of the differential operator on the entire plane $\mathbb{R}^3$ with distributional source yields functions laying in the kernel of the differential operator in the punctured plane. This is ordinarily remedied by implicitly applying a boundary condition; the conformal factor is decomposed into a singular part and a regular part, and the Hamiltonian constraint is extended to the entire plane (\cite{ads:1997PhRvL..78.3606B}), thereby imposing an implicit regularity boundary condition on the remainder while the singular part is chosen arbitrarily to be one of these distributional solutions to the linear part of the differential operator.

To the extent that this ``puncture'' procedure is rigorous, problems of multiplication of distributions can be ignored and the distributions can be used to parameterise the functions lying in the kernel of the differential operator. The resulting functions are regular on the punctured plane and solving for the remainder on the entire plane implicitly chooses a boundary condition by regularity. In this paper, I follow this procedure to produce and study initial data with distributional ``source'' terms.

The layout of the paper is as follows. In \sref{greensFunction} the Green's function for the momentum equation is derived in the conformally flat, maximally sliced transverse-traceless decomposition of the initial value problem for general relativity. It is also shown how the puncture solution to the Hamiltonian constraint arises from a distributional source. In \sref{trumpet} the sources for the trumpet black hole initial data are derived. In \sref{sec:moments} exact initial data having dipole and quadrupole type mass moments are considered while in \sref{sec:nonlinear} spinning trumpet initial data and intial data with higher order momentum moments are considered. Finally, in \sref{discussion}, the results of the paper are summarised. Througout this paper, I use units where $c=G=1$ and indices run over spatial indices unless otherwise specified.

\section{Solutions to the constraint equations parameterised by distributions}
\label{greensFunction}
\subsection{Green's function for the momentum equation}

The Hamiltonian and momentum constraint of Einstein's equations in the ADM formulation are 
\begin{equation}
R + K^2 - K_{ij}K^{ij} = 16\pi\rho
\end{equation}
\begin{equation}
\nabla_j\left(K^{ij} - g^{ij}K\right) = 8\pi j^i
\end{equation}
with $R$ the Ricci scalar of the initial spatial hypersurface with metric $g_{ij}$ and $K_{ij}$ it's extrinsic curvature with trace $K$. When one performs the conformal transverse-traceless decomposition constraint equations and chooses a conformally flat background $g_{ij} = \psi^4 \delta_{ij}$ ($\psi$ is the conformal factor) and maximal slicing $K = 0$, the Hamiltonian and momentum constraint are respectively
\begin{equation}
 \label{eqn:hamConstraint}
 \laplacian\psi + 8^{-1}\psi^{-7} A_{ij} A^{ij} + 2\pi\hat{\rho} = 0,
\end{equation}
\begin{equation}
 \label{eqn:momConstraint}
 \laplacian V_i + 3^{-1} \flatd_i\flatd_j V^j - 8\pi \hat{j}_i = 0,
\end{equation}
where $\hat{\rho}$ and $\hat{j}_i$ are the rescaled energy density and momentum vector respectively given in terms of the normal to the initial hypersurface $n^a$ and the projection tensor $\gamma_{ab}$ and the stress energy tensor as
\begin{equation}
 \hat{\rho} = \psi^5\rho = \psi^5T_{ab}n^an^b
\end{equation}
\begin{equation}
 \hat{j}_i = \psi^6j_i = \psi^6 T_{ab}n^a\gamma^b{}_i
\end{equation}
(in these equations, sums are carried out over the spacetime indices) and $A_{ij} = \flatd_j V_i + \flatd_i V_j - 2/3\delta_{ij}\flatd_kV^k$ is the conformal extrinsic curvature with $K_{ij} = \psi^{-2} A_{ij}$. Initial data satisfying these equations are commonly referred to as Bowen-York initial data. Since the conformal background is flat, indices are raised and lowered with the flat metric and in particular $V_i=V^i$. Because of the use of the flat background, coordinate delta functions will be used in the following analysis.

The basic form of the stress energy tensor of point-like particles as an idealization of extended objects has been derived in a different context in \cite{ads:2010PhRvD..81d4019S}, where neglecting higher order multipole moments, the stress energy tensor of a pointlike source was shown to be
\begin{equation}
  \sqrt{-g}T^{ab} = u^{(a}p^{b)}\delta - S^{c(a}u^{b)}\nabla_c\delta
\end{equation}
where the factor $\sqrt{-g}$ is appearing because of the use of coordinate delta functions. By finding the fundamental solution to the momentum constraint, the source terms for Bowen-York type initial data will be derived and shown to be exactly analogous to these terms.

The Malgrange-Ehrenpreis theorem guarantees the existence of a Green's function for the momentum constraint; to find it, the source is set to $\hat{j}_i(x) = P_i\delta(\bm{x})$ with its Fourier transform being $\tilde{j}_i = P_i$. The fundamental solution is then found by Fourier transforming the momentum constraint, algebraically rearranging terms and integrating to get
\begin{equation}
 \tilde{V}_i = -\frac{1}{4}\left( 7\frac{P_i}{|\bm{x}|}
          + \frac{x_i x^j P_j}{|\bm{x}|^3}\right).
\end{equation}
This is of course the linear momentum solution first derived in \cite{ads:1980PhRvD..21.2047B} showing that the source of linear momentum arises as a point source Dirac delta distribution. This result can immediately be used to write down the general solution to the momentum constraint (with homogeneous boundary conditions at infinity)
\begin{equation}
 \label{eq:momentumSolution}
 {V}_i = -\frac{1}{4}\int\left( 7\frac{\hat{j}_i(\bm{y})}{|\bm{x}-\bm{y}|}
           + \frac{(x_i-y_i) (x^j - y^j) \hat{j}_j(\bm{y})}{|\bm{x}-\bm{y}|^3}\right)d^3{y}.
\end{equation}

This general solution can be used to show that the source $\hat{j}^i = -1/2\epsilon^{ijk}S_k\flatd_j \delta(\bm{x})$ with $\epsilon^{ijk}$ the flat space Levi-Civita tensor gives the vector field
\begin{equation}
 {V}_i = \frac{S^k x^j\epsilon_{ijk}}{|\bm{x}|^3},
\end{equation}
which is precisely the angular momentum solution of Bowen and York.

The combination of the delta function source for the linear momentum term and the gradient term for the angular momentum yields the total source for the momentum equation for puncture initial data
\begin{equation}
 \label{eq:punctureSources}
 \hat{j}_i = \left(P_i - \frac{1}{2}\epsilon_{ijk}S^k\flatd^j \right) \delta(\bm{x}).
\end{equation}

\cite{ads:1980PhRvD..21.2047B} noted that the parameters $P^i$ and $S^i$ are exactly the ADM linear and angular momentum respectively. It is exactly this property that makes the Bowen-York initial data so useful. Heuristically, this can be expected to always be the case on a maximal slice, since the ADM linear and angular momenta (associated with the translational and rotational killing vectors $\xi^a$) as limits of integrals on spheres $\mathcal{S}_r$ can be converted to volume integrals
\begin{equation}
 8\pi P_\xi = \lim_{r\rightarrow\infty}\oint_{\mathcal{S}_r}{K^a{}_b\xi^bdA_a}=\lim_{r\rightarrow\infty}\int_{B_r}{\nabla_a(K^a{}_b\xi^b)dV}
\end{equation}
if one assumes that the extrinsic curvature is valid on the entire open ball $B_r$. The momentum equation \eqref{eqn:momConstraint} can then be used to simplify the volume integrals
\begin{equation}
 P_\xi = \lim_{r\rightarrow\infty} \int_{B_r}{j_b \xi^bdV} = 
\int_{B_r}{j_b\xi^bdV}=\int_{B_r}{\hat{j}_b\xi^bd\hat{V}}
\end{equation}
in the second equality the limit vanishes because of the compact support of $j_i$, $dV$ is the volume form in the initial hypersurface and $d\hat{V}$ being the volume form in the conformal background. When the only source term is a momenta, the translational killing vector measures precisely $P^i$ and when the only source term is a spin the rotational killing vector measures precisely $S^i$. This argument suffers from the weakness of distribution theory in converting from the surface integral to the volume integral and from going from the invariant volume element to the coordinate one. It is completely valid however within the nonlinear generalized function theory of Colombeau algebras (see \cite{grosser2001} for an introduction to this theory).

\subsection{The Hamiltonian constraint}
The usual way for solving the Hamiltonian constraint \eqref{eqn:hamConstraint} in the puncture formalism is by decomposing $\psi = \psi_\mathrm{singular} + u$ and having each component satisfy the following equations respectively
\begin{equation}
 \label{eq:punctureDistributionPart}
 \laplacian \psi_\mathrm{singular} + 2\pi\hat{\rho} = 0,
\end{equation}
\begin{equation}
 \label{eqn:punctureEquation}
 \laplacian u + 8^{-1}\psi^{-7} A_{ij} A^{ij} = 0,
\end{equation}
where \eqref{eq:punctureDistributionPart} is implicitly solved using a pointlike source $\hat{\rho} = M\delta(\bm{x} - \bm{x}_I)$ so that $\psi_\mathrm{singular} = 1 + M/2|\bm{x}-\bm{x}_I|$. Then one needs to solve only the single equation (usually numerically) for puncture initial data, \eqref{eqn:punctureEquation}. For vanishing extrinsic curvature, it is already known from \cite{ads:1979JMP....20.1423P} that the source term for Schwarzschild solution is a delta function, although there it is associated with a physical singularity while here it is associated with a coordinate singularity.

One should note that the original problem is ill posed (the solution is not unique) on the punctured plane $\mathbb{R}^3 - \{\bm{x}_I\}$ without suitable boundary conditions at the origin. Namely, any solution to \eqref{eq:punctureDistributionPart} having distributional source $\hat{\rho}$ at the punctures lies in the kernel of the Laplacian; for instance the mass term $M$ above is not unique--any mass will solve the Hamiltonian constraint \eqref{eqn:hamConstraint}. This ill-posedness could be removed by considering the original solution on the entire plane and considering the confomal factor to have a Dirac delta source term, in this case the puncture formalism is just a form of regularization but a nonlinear theory of distributions is necessary. Alternately, the above decomposition and implied regularity on $u$ can be considered in lieu of a boundary condition. This ill-posedness is often exploited to iterate over the bare mass until the black hole has a desired horizon mass as outlined in \cite{ads:2008PhRvD..77b4027B}. Parameterising the singular part via a distribution is preferable to appealing to a topological condition such as a compactified alternate world sheet if only for its simplicity.

\subsection{Trumpet initial data}
\label{trumpet}
It was numerically verified in \cite{ads:2007JPhCS..66a2047H} that the evolution under dynamic gauge conditions of puncture spacetimes causes the power of the singularity of the conformal factor to change in time. For this reason, the exact slicing of the Schwarzschild geometry that satisfies a gauge condition close to the normal gauge condition used in evolution was found by \cite{ads:2007PhRvD..75f7502B}. This solution has been used to produce initial data in \cite{ads:2009PhRvD..80f1501I,ads:2009PhRvD..80l4007H} on the punctured plane which has been termed ``trumpet'' initial data. It should be emphasized that with zero momentum the trumpet black hole is identically a Schwarzschild black hole. Since the formulation used is again Bowen and York's conformally flat, transverse traceless decomposition, the singular structure which is implictly parameterising the initial data can be studied using the Green's function from the previous section.

For a trumpet black hole, there is a background extrinsic curvature in addition to the Bowen-York spin and momentum terms. This background extrinsic curvature is given by \cite{ads:2009PhRvD..80f1501I}
\begin{equation}
A_{ij} = \frac{3\sqrt{3}M^2}{4r^3}\left(  \delta_{ij} - \frac{x_ix_j}{r^2} \right)
\label{eq:trumpetA}
\end{equation}
and the corresponding exact conformal factor is
\begin{equation}
 \psi = \sqrt{\frac{R(r)}{r}}
\end{equation}
with the areal radius $R(r)$ given implicitly in terms of the isotropic radius r
\begin{equation}
\begin{array}{rcl}
 r &=& \displaystyle{\left[\frac{2R + M + \sqrt{4R^2 + 4MR + 3M^2}}{4} \right]}
    \displaystyle{\times\left[\frac{(4+3\sqrt{2})(2R-3M)}{8R + 6M + 3\sqrt{8R^2 + 8MR + 6M^2}}\right]^{1/\sqrt{2}}}.
\end{array}
\end{equation}

The presence of terms in the extrinsic curvature which scale inversely as $r^3$ is an indication that the momentum sources must contain derivatives of the delta function. A generic source of the form
\begin{equation}
 \hat{j}_i = B_i^j\flatd_j\delta(\bm{x})
\end{equation}
together with the general solution to the momentum constraint \eqref{eq:momentumSolution} can be used to determine the corresponding first order solution to the momentum constraint. Coefficients of $x^ny^mz^p$ are collected to determine the unknown coefficients $B_i^j$ such that \eqref{eq:trumpetA} is reproduced by this source term. In this way, the trumpet momentum source is determined to be
\begin{equation}
 \hat{j}_i = \frac{\sqrt{3}M^2}{4}\flatd_i \delta(\bm{x} - \bm{x}_I).
\end{equation}

Interestingly, the trumpet initial data has no source for the Hamiltonian constraint. Integrating the Hamiltonian constraint against some test function $\varphi$, one has,
\begin{equation}
-\int 2\pi \hat{\rho}\varphi d\hat{V} = \int{ \left(\psi\laplacian\varphi + \frac{A_{ij}A^{ij}}{8\psi^7}\varphi\right) d\hat{V}}
\end{equation}
where $\varphi$ is a compactly supported test function with support at the location of the singularity of $\psi$ and $d\hat{V}$ is the flat volume element of the background metric. This integral is converted to a limit
\begin{equation}
 -\int 2\pi \hat{\rho}\varphi d\hat{V} = \lim_{r\rightarrow 0} \int_{\mathbb{R}^3-B_r}{ \left(\psi\laplacian\varphi + \frac{A_{ij}A^{ij}}{8\psi^7}\varphi\right) d\hat{V}}
\end{equation}
and integrated by parts
\begin{equation}
\begin{array}{rcl}
 -\int 2\pi \hat{\rho}\varphi d\hat{V} &=& \displaystyle{\lim_{r\rightarrow 0} \int_{\mathbb{R}^3-B_r}{ \left(-\flatd_i\psi\flatd^i\varphi + \frac{A_{ij}A^{ij}}{8\psi^7}\varphi\right) d\hat{V}}}
\displaystyle{+ \lim_{r\rightarrow 0}\oint_{\mathcal{S}_r}{\psi\frac{d\varphi}{dr}}dS}
\end{array}
\end{equation}
with $dS$ the area element over the surface of the open ball $\mathcal{S}_r$. Since the radial derivative of the test function is uniformly bounded, the surface integral is limited as
\begin{equation}
 \left|\lim_{r\rightarrow 0}\oint_{\mathcal{S}_r}{\psi\frac{d\varphi}{dr}}dS\right| \leq \sup{\left|\frac{d\varphi}{dr}\right|}\lim_{r\rightarrow 0}\oint_{\mathcal{S}_r}{\psi}dS = 0
\end{equation}
with the equality arising from computation with the conformal factor. One more integration by parts yields
\begin{equation}
\begin{array}{rcl}
 -\int 2\pi \hat{\rho}\varphi d\hat{V} &=& \displaystyle{\lim_{r\rightarrow 0} \int_{\mathbb{R}^3-B_r}{ \left(\laplacian\psi + \frac{A_{ij}A^{ij}}{8\psi^7}\right)\varphi d\hat{V}}}
\displaystyle{- \lim_{r\rightarrow 0}\oint_{\mathcal{S}_r}{\varphi\frac{d\psi}{dr}}dS} = 0.
\end{array}
\end{equation}
where the volume integral vanishes due to the momentum constraint (there are no sources outside of the open ball) and the surface integral vanishes by an analogous uniform bounding as above.

In the case of the trumpet data, the source of the mass of the spacetime arises entirely out of an isotropic ``momentum'' source. For this reason, separating the Hamiltonian constraint to factor out the singular part of the conformal factor is necessary for trumpet data only to formulate a numerical method that does not have a singular solution whereas for the original puncture data it is necessary to deal with a fundamental ill posedness and select a mass term. In this case, where one factors out the singularity only for numerical purposes, it suffices to factor out only the singularity $\sqrt{3M/2r}$ in order to ensure that the remainder $u$ is bounded (although a more sophisticated factor is needed for spinning trumpets), and the Hamiltonian constraint is well defined over all $\mathbb{R}^3$. This fact is used in \sref{sec:nonlinear} to produce spinning trumpet initial data as well as more general initial data.
\section{Initial data with higher order mass moments}
\label{sec:moments}
Far from a source, the source itself may be idealized as a point particle with various mass moments (usually described in terms of multipoles). In this case the only equation to be solved is the linear part of the Hamiltonian constraint \eqref{eqn:hamConstraint}. The matter distribution that enters the field equations in this case is a series of derivatives of the delta function. In Newtonian physics, where mass is always positive, a dipole moment (the first order moment) is always zero except when the first order moment is computed about an origin which is not the barycentre; it is unclear to what extent this is meaningful in general relativity. A quadrupole moment (the trace-free second order moment) quantifies the asymmetry of the distribution. Considering the first two matter moments, the conformal source term including up to quadrupole terms is
\begin{equation}
  \label{eq:momentDistribution}
 \hat{\rho} = \left(1 - D^i\flatd_i + R^{ij}\flatd_i\flatd_j \right) \delta (\bm{x})
\end{equation}
with $D^i$ quantifying the dipole moment and $R^{ij}$ quantifying the second order moment (not the trace-free quadrupole moment). For a generic matter distribution with unit mass, the terms in this idealization can be computed as $D^i = \int{\hat{\rho}x^id\hat{v}}$ and $R^{ij}=\int{\hat{\rho}x^ix^jd\hat{v}}$. It is possible to derive many forms for the second order moment, however for a ring with negligible cross section
\begin{equation}
 R^{ij} = \frac{q}{4}(\delta^i_{x}\delta^j_{x} + \delta^i_{y}\delta^j_{y})
\end{equation}
with scalar quadrupole moment $q$ being a combination of the mass and radius of the ring. If the dipole moment is aligned to the $z$ axis so that $D^i = (0,0,d)$, the solution to the Hamiltonian constraint \eqref{eqn:hamConstraint} with source \eqref{eq:momentDistribution} is given by
\begin{equation}
 \psi = 1 + \frac{1}{2r} + \frac{dz}{2r^3} + \frac{q}{8r^5}(x^2 + y^2 - 2z^2)
\end{equation}
so long as $\hat{j}_i = 0$ and so $A_{ij} = K_{ij} = 0$.

\begin{figure}[ht]
  \begin{center}
    \includegraphics[width=0.5\columnwidth]{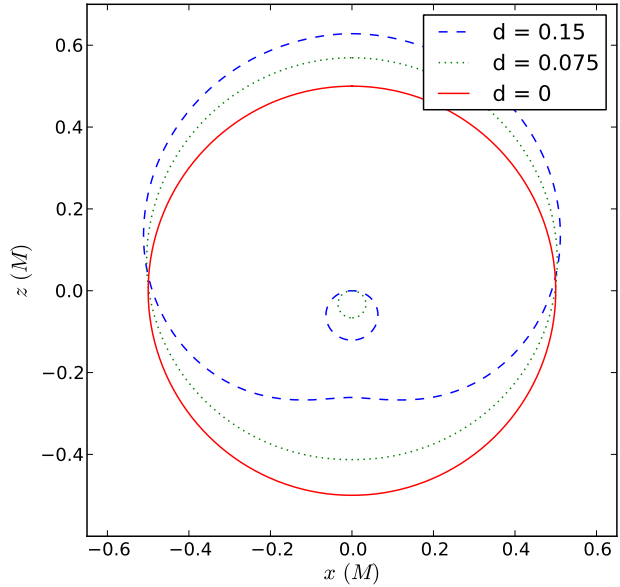}
    \caption{\label{fig:dipoleHorizons} Apparent horizons (outer curves) and
	coordinate singularities (inner curves) plotted
	for spacetimes with various dipole moments $d$ added.
	The inner curves represent coordinate singularities in the 
	metric where the spatial metric vanishes. The centroid of the apparent
	horizon shifts upwards with increasing dipole moment while the centroid
	of the coordinate singularity shifts downward.}
  \end{center}
\end{figure}

\begin{figure}[ht]
  \begin{center}
    \includegraphics[width=0.5\columnwidth]{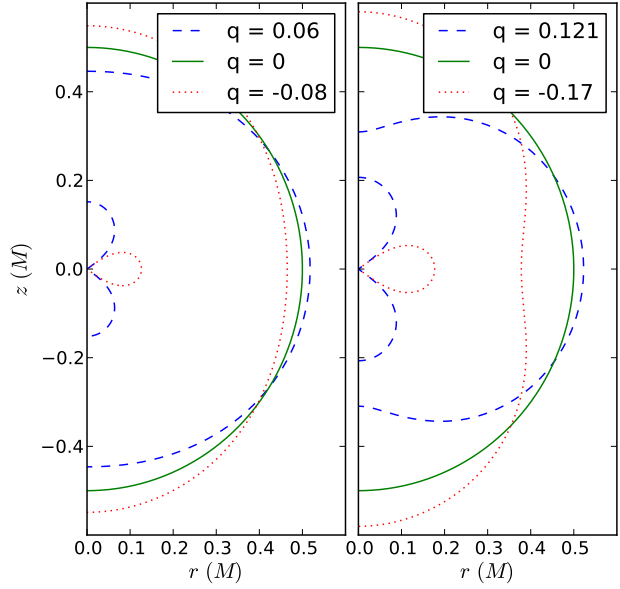}
    \caption{\label{fig:quadrupoleHorizons} Apparent horizons (outer curves) and
	coordinate singularities (inner curves) plotted
	for spacetimes with various quadrupole moments $q$ added. For small
	deformations, as seen in the left plot, the horizons are
	nearly elliptical. For larger deformations, as seen in the right plot, indentations occur
	on the horizon. The inner curves represent coordinate singularities in the 
	metric where the spatial metric vanishes.}
  \end{center}
\end{figure}

These intitial data generically contain coordinate singularities with spherical topology: a closed surface where the conformal factor goes to zero. In the physical space, these surfaces would correspond to extended points; all points on the coordinate surface are separated by zero distance.  As seen in \fref{fig:dipoleHorizons}~and~\fref{fig:quadrupoleHorizons} these coordinate singularity surfaces are generically contained within the apparent horizons of the spacetimes (see \cite{ads:1996PhRvD..54.4899T} for an overview of apparent horizons and horizon finding). The interior regions of these coordinate singularities are topologically $\mathcal{S}^3$. Perhaps more interestingly, these surfaces of equidistant points ensure that the origin is a finite distance away from any other point in the plane; much like the trumpet initial data, it is no longer possible to interpret the origin to be the conformally compactified infinity of an alternate world sheet.

The qualitative effect of adding the multipole moments is to change the apparent horizon in a corresponding fashion. The dipole moment, for small values of $d$ shifts the horizon in the respective direction while maintaining an overall spherical shape, as seen in \fref{fig:dipoleHorizons}. For larger values of the dipole moment, the apparent horizon becomes dimpled in the direction $-D^i$. When the value of the quadrupole moment is small, the overall shape of the apparent horizon is elliptical, as seen in the left plot of \fref{fig:quadrupoleHorizons}. For large magnitude quadrupole moments, the apparent horizon becomes indented on the equator for negative moments, and at the poles for positive moments.

\begin{figure}[ht]
  \begin{center}
    \includegraphics[width=0.5\columnwidth]{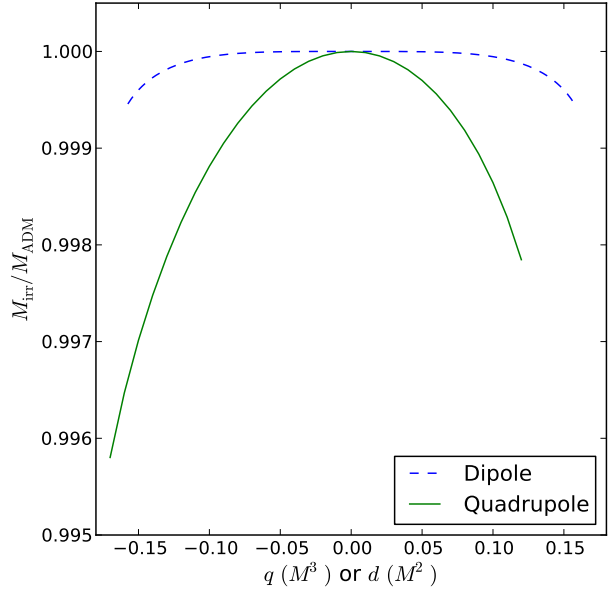}
    \caption{\label{fig:momentMass} Irreducible mass of non-spinning, zero momentum
    black hole with added dipole and quadrupole moments. The difference between
    the irreducible mass and the ADM mass $M_\mathrm{ADM} = 1$ estimates the fractional 
    radiation content of these spacetimes. Spacetimes with dipole moments have 
    very little radiation content in comparison to the spacetimes with quadrupole 
    moments. The extent of the curves shows the range for which apparent horizons 
    could still be numerically found numerically.}
  \end{center}
\end{figure}

These deformations induced on the horizons by the added structure of the distributional source terms cause the horizon mass of the black hole to decrease with increasing magnitude of the moment. The mass of the black hole is identified with its irreducible mass since there is no angular momentum,
\begin{equation}
M_\mathrm{irr} = \sqrt{\frac{A}{16\pi}}.
\end{equation}
The masses of the black holes considered in this section are plotted in \fref{fig:momentMass}. The difference between irreducible mass $M_\mathrm{irr}$ and ADM mass is an estimate for the total amount of radiation in the spacetime. From the figure, one sees that for the dipole black holes, where the horizon could be found, the maximum amount of radiation is less than $0.1\%$. The horizon mass of the black holes with quadrupole moments is symmetric with respect to the sign of the quadrupole moment, which is unexpected given the difference in the structure of the apparent horizon for positive and negative quadrupole moments. For the horizons found, the maximum radiation content is about $0.2\%$ and $0.5\%$ for positive and negative quadrupole moments respectively.

\section{Fully non-linear initial data with distributional sources}
\label{sec:nonlinear}
\begin{figure}%
	\begin{center}
	\includegraphics[width=0.5\columnwidth]{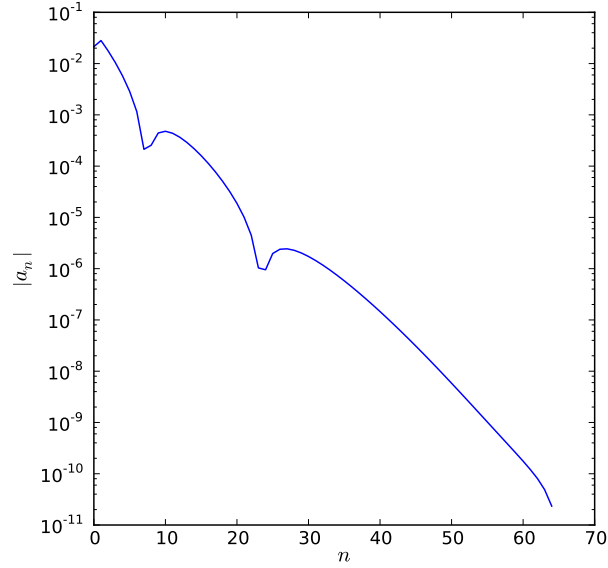}
		\caption{The truncation error for a spinning puncture solution in spherical coordinates with $J=0.5$ and $M=1$. The exponential fall-off with linear increase in basis order is indicative of the function having an exponential index of convergence. The use of spherical coordinates is crucial to achieve this result.}%
		\label{fig:punctureTruncation}%
	\end{center}
\end{figure}

Considering completely general sources for the extrinsic curvature necessitates solving the fully non-linear Hamiltonian equation. Considering derivatives of the Dirac delta to higher than second order will lead to stronger singularities appearing in the conformal factor--the $m/2r$ term is insufficient to ensure regularity. This also occurs if one tries to solve the Hamiltonian equation for the bare trumpet conformal factor numerically. Furthermore, for spinning trumpet initial data, the exact implicit conformal factor is not strong enough to make the solution fully regular since the spin term (an antisymmetric derivative) enters at the same order as the mass term (a gradient). For numerical regularity for general sources, the following decomposition of the conformal factor is proposed:
\begin{equation}
\psi = 1 + \psi_\rho + \frac{\sigma}{r^p} + u.
\label{eq:regularConformal}
\end{equation}
Here $\psi_\rho$ satisfies the Poisson equation $\laplacian \psi_\rho +2\pi\hat{\rho} = 0$, the term $\sigma$ depends only on angle, and $u$ is the bounded remainder part of the conformal factor. The extrinsic curvature term $A_{ij}A^{ij}$ can always be written in terms of a denominator which is a power of radius $r$ and a numerator which is polynomial in $r$ and trigonometric functions of the angles. Considering only axisymmetric sources and using spherical coordinates, the extrinsic curvature is decomposed as
\begin{equation}
A_{ij}A^{ij} \equiv \frac{\kappa(r,\theta)}{r^n} \equiv \frac{\sum_{i=1}^{n-1}r^i\alpha(\theta) + \beta(\theta)}{r^n}
\label{eq:decomposedASquared}
\end{equation}
with $n$ the problem dependent integer such that $\beta\neq 0$. By demanding that $\displaystyle{\lim_{r\rightarrow 0}r^{p+2}\laplacian u = 0}$, one concludes that $8p + 2 - n = 0$ and the Hamiltonian constraint yields the following two equations 
\begin{equation}
\triangle_\Omega \sigma + p(p-1)\sigma +\frac{\beta}{8\left(\sigma + \displaystyle{\lim_{r\rightarrow 0}r^p\psi_\rho}\right)^7}=0,
\label{eq:angularPart}
\end{equation}
\begin{equation}
r^{p+2}\laplacian u - \frac{\beta}{8\left(\sigma + \displaystyle{\lim_{r\rightarrow 0}r^p\psi_\rho}\right)^7} +
	\frac{1}{8}\frac{\kappa}{\left(r^p\left(1+u+\psi_\rho\right) + \sigma\right)^7} = 0
\label{eq:regularHamiltonian}
\end{equation}
with $\triangle_\Omega = \frac{\partial^2}{\partial\theta^2} + \cot\theta\frac{\partial}{\partial\theta}$ the angular part of the Laplacian multiplied by the radius squared. These equations would be analagous for non axisymmetric sources except that $\triangle_\Omega$ would be replaced by the full angular part of the Laplacian. To study solutions to these equations, Chebyshev pseudo-spectral methods are used. In particular, the collocation points are chosen as the roots of the Chebyshev polynomials in order to ensure regularity at the origin and symmetry axis without the use of boundary conditions. In spherical coordinates, the ordinarily $\mathcal{C}^2$ puncture remainder is much more regular and exponential convergence can be achieved as noted in  \cite{ads:2004PhRvD..70f4011A}. The exponential convergence is shown in \fref{fig:punctureTruncation} where the truncation error of a spinning puncture dataset displays an exponential index of convergence.

\subsection{Spinning trumpet initial data}
\begin{figure}%
	\begin{center}
	\includegraphics[width=0.5\columnwidth]{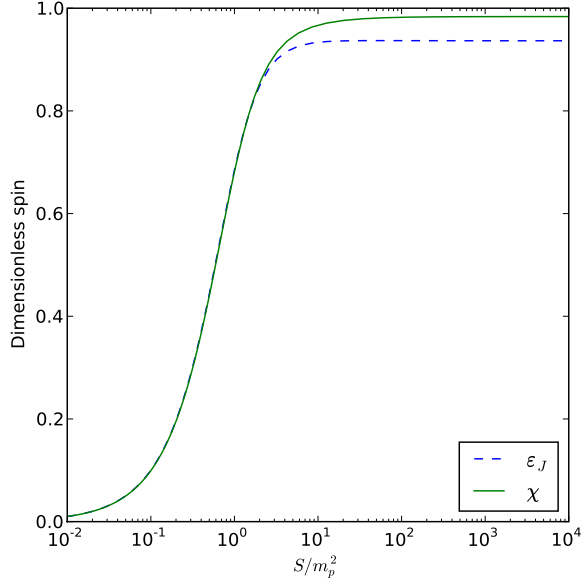}%
	\caption{The dimensionless spin of the black hole measured on the horizon $\chi$ and at infinity $\varepsilon_J$ as a function of  bare dimensionless spin. Although the black hole has an initially large value of the dimensionless spin, it accretes gravitational radiation during evolution which increases its mass and subsequently decreases its dimensionless spin. This process causes the black hole to relax, under evolution, to a dimensionless spin close to the ADM value.}%
	\label{fig:trumpetSpins}%
	\end{center}
\end{figure}

In the case of non-spinning trumpet initial data (with or without momentum) $\beta = 81/8 M^4$ which leads $p=1/2$ and $\sigma = \sqrt{3M/2}$ so that the singular part of the conformal factor is $\psi_s = 1 + \sqrt{3M/2r}$. For spinning trumpets, it is insufficient to subtract this factor or even the exact trumpet conformal factor as shown in \cite{ads:2009PhRvD..80f1501I}. However, it is possible to produce this initial data with the above decomposition where the coefficient of the singular part $\sigma$ has angular dependence. In the case of spinning trumpet initial data the extrinsic curvature is generated from the general solution \eqref{eq:momentumSolution} using the source
\begin{equation}
 \widehat{j}_i = \left(S_z\epsilon_{ijz}\partial^j + \frac{\sqrt{3}M^2}{4} \partial_i\right) \delta(x).
\end{equation}
Due to the fact that the limiting case of both trumpet and puncture initial data with spins will have the same source term $\widehat{j}_i = S_z\epsilon_{ijz}\partial_j \delta(x)$ the limiting behaviour of the solutions are identical. However, the transition from the puncture type $1/r$ singularity to the spinning trumpet type $1/\sqrt{r}$ singularity is singular for punctures while the transition is smooth for the trumpet initial data produced by the method here. The exact singular solution may even be computed. It is known for Bowen-York type initial data that it is not possible to produce black holes with arbitrarily large ratios of spin to mass; this was detailed in \cite{ads:2008PhRvD..78h4017L}. To quantify the behaviour of the spinning trumpets, the dimensionless spin of the black hole
\begin{equation}
\chi = \frac{S_z}{M^2}
\label{eq:chi}
\end{equation}
is monitored at the initial hypersurface using a Christodoulou-like mass 
\begin{equation}
M^2 = M_\mathrm{irr}^2 + \frac{S_z^2}{M_\mathrm{irr}^4}.
\label{eq:christodoulouMass}
\end{equation}
Due to the axisymmetry of the system, the spin measured on the horizon is identical to the ADM spin which is identical to the parameter $S_z$ used to construct the extrinsic curvature. During evolution, the spin measured on the horizon will quickly relax to a lower equilibrium value and an excellent approximation of the relaxed value is the dimensionless spin as measured at infinity as argued by \cite{ads:2008PhRvD..78h4017L}
\begin{equation}
\varepsilon_J = \frac{S_z}{M_\mathrm{ADM}^2}
\label{eq:relaxedSpin}
\end{equation}
here $S_z$ is a parameter of the problem but all of the masses must be measured. These quantities are displayed in \fref{fig:trumpetSpins} where the bare parameters have been varied by many orders of magnitude. Exactly like the spinning puncture initial data, it is not possible to produce black holes of arbitrary spin parameter. Although the total angular momentum of the solution can be set to an arbitrary value, the mass of the spacetime increases yielding an asymptotic maximum value of the dimensionless spin. One sees that the limiting value of the dimensionless spin is $\chi \approx 0.984$ which will relax upon evolution to approximately the asymptotic value of $\varepsilon_J \approx 0.937$. The data presented here has been run at multiple resolutions to ensure accuracy to one part in one million, the data presented are results run with the number of collocation points set to $(n_r, n_\theta) = (80,40)$.

\subsection{Higher order momentum moments}
In a close binary, it is likely that the black holes will locally have a Kerr like geometry at first order, but due to tidal effects will have induced moments on their horizons. It may be desirable to incorporate these effects in the source terms used to construct initial data (or in the shape of apparent horizon). For this reason, in this section black holes with higher order momentum moments are considered. In particular, the following source is considered for the momentum constraint
\begin{equation}
\hat{j}_i = -\left( \frac{1}{4}\epsilon_{ijz}\flatd^j + Q_z\epsilon_{ijz}\flatd^j\flatd_z + Q_c\epsilon_{ijk}\epsilon^{jzm}\flatd_m\flatd^j\right) \delta(\bm{x})
\label{eq:momentumQuadrupole}
\end{equation}
with the factor $1/4$ chosen to have a high but not numerically problematic spin $\chi_0\approx0.423$. The two terms are respectively an axial derivative of the spin term (proportional to $Q_z$) and the curl of the spin term (proportional to $Q_c$), both give rise to axisymmetric systems angular dependence of a higher frequency than the spin term. The Hamiltonian constraint is chosen to have source $\hat{\rho} = \delta(\bm{x})$ to make the solutions as regular as possible. Due to the increased irregularity of the solutions, $(n_r,n_\theta)=(120,50)$ collocation points were used to achieve convergence over the range of parameters studied here. 

\begin{figure}%
	\begin{center}
	\includegraphics[width=0.5\columnwidth]{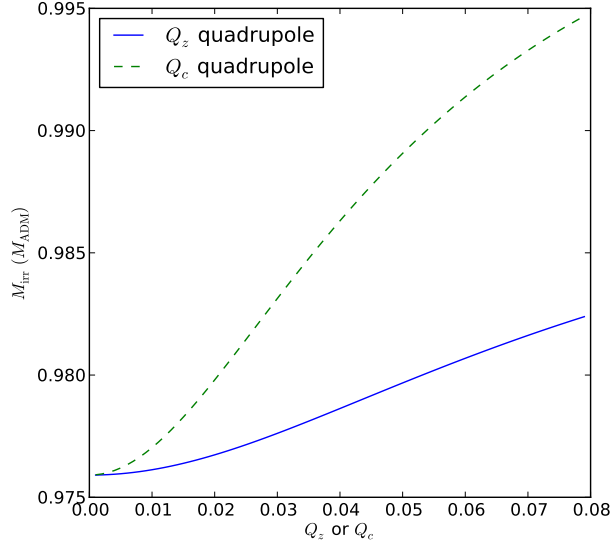}
		\caption{The change in mass for a spinning puncture as the momentum quadrupole moments are increased. The irreducible mass is seen to increase with the quadrupole moment, from this one can conclude that the fractional radiation content in the spacetime is decreasing.}%
		\label{fig:momentumMomentMass}%
	\end{center}
\end{figure}
\begin{figure}%
	\begin{center}
	\includegraphics[width=0.5\columnwidth]{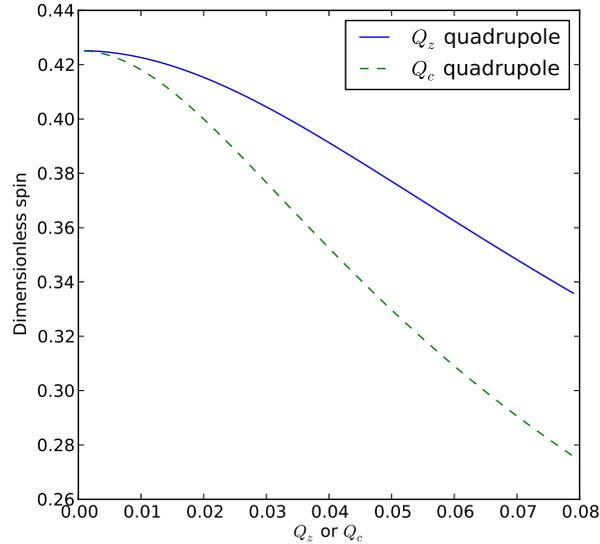}
		\caption{The change in dimensionless spin for a spinning puncture as the momentum quadrupole moments are increased. Due to the increase in mass as seen in \fref{fig:momentumMomentMass} the dimensionless spin decreases.}%
		\label{fig:momentumMomentSpin}%
	\end{center}
\end{figure}

The effects of the quadrupole moments on the mass and spin of the black hole are plotted respectively in \fref{fig:momentumMomentMass} and \fref{fig:momentumMomentSpin} respectively. The Christodoulou-like mass is not plotted for these spacetimes, as the black holes are no longer even approximately Kerr-like. It is seen from \fref{fig:momentumMomentMass} that the irreducible mass of the black hole increases with increasing quadrupole moment. Since the irreducible mass can never increase over the value of the ADM mass (this is the Penrose inequality, proved by \cite{ads:2001JDGeo..59..177B}) during evolution, there must be less radiation in the spacetime. Because of the increase in the irreducible mass of the black hole, the dimensionless spin of the black hole must decrease. This effect is shown in \fref{fig:momentumMomentSpin}, where a very steep decrease in the spin is seen with an increase in the momentum quadrupole. This effect could have consequences for spinning black holes in a binary system; if the black holes are tidally interacting, the induced quadrupole moment could prevent the black hole from having an extremal spin by increasing its irreducible mass and with it enforcing cosmic censorship.
\section{Discussion}
\label{discussion}
The above analysis revealed that the puncture (and trumpet) initial data had as its origin distributional source terms. Two distinct terms that gave rise to mass terms were found, a delta function source for the Hamiltonian constraint in the case of puncture initial data and an isotropic delta function gradient source for the momentum constraints in the trumpet initial data. The linear momentum of the puncture initial data, and subsequently the trumpet initial data, arises from a delta function source while the angular momentum arises from the contraction of an anti-symmetric (spin) tensor with the gradient of a delta function.

It is possible to extend the method (parameterisation by distributions of the vacuum solutions) to the conformal thin sandwich formulation or to use alternate conformal backgrounds. The basic approach being to parameterise solutions lying in the kernel of the linear part of the differential equations, then to decompose the conformal factor into a source, singular and regular part and finally to solve for the regularised remainder on the entire plane. This process was used to generate novel black hole initial data in \sref{sec:nonlinear}. Given the heuristic argument presented in \sref{greensFunction}, the same source terms would need to be used to generate the extrinsic curvature for linear and angular momentum. It is possible to put this heuristic argument on rigorous footing using Colombeau algebras but one must then tackle the issue of uniqueness with respect to mollifier if using special algebras.

It is also possible to extend these results to the production of binary black hole initial data. For binary data, especially binary trumpet data, it may be preferable to produce individual spinning, boosted trumpets using the methods above to high order and then subtract the combined remainders so that the singular parts are correct to next to leading order. The resulting remainder should be much more regular in this way. The higher order moments used here may become important in a binary, where tidal effects will distort the horizon (\cite{ads:2010PhRvD..81b4029P}) and with it possibly the sources.

The code used in the production of the nonlinear initial data studied here is freely available under the Gnu General Public License at https://github.com/SwampWalker/LeapingMonkey. Readers interested in extending these results are welcome and encouraged to build upon the code used here.
\ack
I would like to thank Juan Barranco, Barry Wardell, Ian Hinder and Abraham Harte for their useful discussions and insight. Partial support comes from the DFG Grant SFB/Transregio 7.
\bibliographystyle{jphysicsB}
\bibliography{punctureSources}
\end{document}